\documentclass[]{aa}
\usepackage{txfonts}
\usepackage{graphicx}
\usepackage{hyperref}

\usepackage[dvipsnames]{xcolor}

\usepackage[utf8]{inputenc}
\usepackage{comment}
\usepackage{siunitx}

\newcommand{\DHA}[1]{{{#1}}}

\begin{document}

\title{\emph{Fermi} Large Area Telescope Observations of the Fast-dimming Crab Nebula in 60--600 MeV}

\subtitle{}

\titlerunning{\emph{Fermi} LAT Observations of the Fast-dimming Crab Nebula}

\author{Paul K. H. Yeung\inst{1}
        \and
        Dieter Horns\inst{1}}

\institute{Institute for Experimental Physics, Universität Hamburg, Luruper Chaussee 149, D-22761 Hamburg, Germany\\
\email{kin.hang.yeung@desy.de} }

   \date{Received September 19, 2019; accepted May 5, 2020}

\authorrunning{P. K. H. Yeung \& D. Horns}

\abstract
{The Crab pulsar and its nebula are the origin of 
relativistic electrons which can be observed through their
synchrotron and inverse Compton emission. The transition between
synchrotron-dominated and inverse-Compton-dominated emissions takes place  at $\approx 10^9$~eV.}
{The short-term (weeks to months) 
flux variability of the synchrotron emission from the most energetic electrons is investigated with data from ten 
years of observations with the \textit{Fermi} Large Area Telescope (LAT) in the energy range from 60~MeV to 600 MeV.}
{The off-pulse light-curve has been reconstructed from
phase-resolved data. The corresponding histogram of flux measurements is used to identify distributions of flux-states and the statistical significance of a lower-flux component is estimated with dedicated simulations of mock light-curves. The energy spectra for different flux states are reconstructed.}
{We confirm the presence of flaring-states which follow a log-normal flux distribution. Additionally, we discover
a low-flux state where the flux drops to as low as 18.4\%
of the intermediate-state average flux and stays there for several weeks.
The transition time is observed to be as short as 2  days. The energy spectrum during the
low-flux state resembles the extrapolation of the inverse-Compton spectrum measured at energies beyond several GeV energy, \DHA{implying that the high-energy part of the synchrotron emission is dramatically depressed}.}
{The low-flux state  found here and the transition time of at most 10 days indicate  that the bulk ($>75$\%) of the synchrotron emission above $10^8$~eV originates in a compact volume with apparent angular size of $\theta\approx\ang[angle-symbol-over-decimal]{;;0.4}~t_\mathrm{var}/(5~\mathrm{d})$. We tentatively infer that the so-called inner knot feature is the origin of the  bulk of the $\gamma$-ray emission. }

\keywords{}

\maketitle


\section{Introduction} 
\label{sec:intro}
Isolated neutron stars and their environments are powerful sites of particle acceleration,
which result in the formation of pulsar wind nebula (PWN) systems. 
In case of the Crab Nebula, the extended cloud of non-thermal plasma is radiating in multi-wavelength, from radio to
gamma-ray \citep{aharonian_et_al._crab_2004,buhler_surprising_2014,dubner_morphological_2017}.

The Crab Nebula is a PWN powered by a $\sim$1 kyr old pulsar \citep{hester_crab_2008}. 
It is a part of the core-collapse supernova remnant located in the constellation of Taurus
and at a distance of 2 kpc \citep{trimble_motions_1968}.
Due to the exceptionally wide observable energy range, we can study the processes of particle acceleration presumably happening at the termination shock and witness energy-losses in the nebula \citep[e.g.,][]{spitkovsky_time_2004,fraschetti_particle_2017}.

The observed hard $\gamma$-ray (1~GeV--80~TeV) spectrum of the Crab Nebula  has been compared with various 
model calculations which use widely different approaches \citep{de_jager_expected_1992,atoyan_mechanisms_1996,hillas_spectrum_1998,volpi_non-thermal_2008,meyer_crab_2010,martin_time-dependent_2012}.
All these models assume that the gamma-ray emission in this energy range is predominantly produced via inverse-Compton (IC) scattering of relativistic electrons with synchrotron-radiated photons as initially suggested by \citet{rees_implications_1971} and \citet{gunn_motion_1971}. 

Meanwhile, at lower energies, the observed nebular 
 $\gamma$-ray (0.75~MeV--1~GeV) spectrum is presumably dominated  
by the synchrotron mechanism \citep[][]{Kuiper_2001, buehler_gamma-ray_2012}. 
The Crab Nebula experiences recurrent flares (roughly one per year) detected with AGILE and \emph{Fermi} Large Area Telescope (LAT), 
some of which boosted up the $>$100 MeV synchrotron flux by a factor of $\gtrsim20$
\citep[e.g., ][]{Tavani_Discovery_2011, Abdo_Gamma-Ray_2011, buehler_gamma-ray_2012, mayer_rapid_2013}. Enhanced $\gamma$-ray emission of the synchrotron component can last for a broad variety of timescales ranging from days to weeks \citep{Striani2013}. Ongoing instability of the synchrotron emission from the Crab Nebula is also observed in the hard X-ray/soft $\gamma$-ray regime over a longer range of time \citep{Ling_Wheaton_2003, Wilson-Hodge2011}.

In this work, we study the $\gamma$-ray variability of the Crab Nebula  in detail, 
with the  $>$60~MeV LAT data accumulated over $\sim$10 years during the off-pulse phase of the Crab pulsar. 
In addition to the flaring periods, we consider the entire light-curve. 

\section{Data reduction \& analysis} \label{sec:data}
We perform a series of binned maximum-likelihood analyses (with an angular bin size of $0.1^\circ$) for a region of
interest (ROI) of $30^\circ\times30^\circ$  centered at RA=$05^{h}34^{m}31.94^{s}$,
Dec=$+22^\circ00'52.2"$ (J2000), which is approximately the center of the Crab
Nebula \citep{lobanov_vlbi_2011}. 
We  use the data of 60~MeV--10~GeV photon energies,
registered with the LAT between 2008 August 4 and 2018 August 20. The data are reduced
and analyzed with the aid of the \emph{Fermi} Science Tools v11r5p3 package.

Considering that the Crab Nebula is quite close to the Galactic plane (with a
Galactic latitude of $-5.7844^\circ$), we adopt the events classified as
Pass8 ``Clean" class for the analysis so as to better suppress the background.
The corresponding instrument response function (IRF) ``P8R2$_-$CLEAN$_-$V6" is
used throughout the investigation. Only the data collected during the off-pulse phase 
(0.56-0.88; we adopt the same convention of phase as in \citet{buehler_gamma-ray_2012}) of the Crab pulsar are selected for analysis. 
Correspondingly, a correction factor of 1/0.32 is taken into account in calculations of phase-averaged fluxes. We further filter the data by accepting
only the good time intervals where the ROI was observed at a zenith angle less
than 90$^\circ$ so as to reduce the contamination from the albedo of Earth.

In order to account for the contribution of diffuse background emission, 
we  include the Galactic 
background (gll$_-$iem$_-$v06.fits), the isotropic background 
(iso$_-$P8R2$_-$CLEAN$_-$V6$_-$v06.txt) as well as all other  point sources
cataloged in the LAT 8-year Point Source Catalog \citep[4FGL;][]{Fermi_Fourth_2019}
within 32$^\circ$ from the ROI center in the source model.  We  set free the
spectral parameters of the  sources within 10$^\circ$ from the ROI center (including the prefactor and index of the Galactic 
diffuse background as well as the normalization of the isotropic background) in the
analysis. For the sources at angular separation beyond 10$^\circ$ from the ROI center, their
spectral parameters are fixed to the catalog values. 

The three point sources located within the nebula are cataloged 
as  4FGL J0534.5+2200, 4FGL J0534.5+2201i, and 4FGL J0534.5+2201s, which respectively model the
Crab pulsar, the IC, and synchrotron components of the Crab Nebula. 
We remove 4FGL J0534.5+2200 from the source model because the on-pulse data has been screened out.

In broadband spectral analyses, we enable the energy dispersion correction which operates on the count spectra of most sources including the Crab Nebula, following the recommendations of the \emph{Fermi} Science Support Center.

\section{Spectral properties \& variability of the Crab Nebula}

\subsection{Time-averaged spectrum} \label{subsec:TimeAveSpec}

The energy spectrum of the off-pulse nebular component at energies between 60~MeV and 10~GeV
is reconstructed using the  combined observational data of approximately ten years 
(see previous section for an overview of the data reduction steps including the pulsar gating).

The data are fit by a two-component (additive) model. 
Similar to a previous study \citep{buehler_gamma-ray_2012}, we use the superposition of a soft power law (PL) 
with a photon index constrained to the interval $3$-$5$ for the synchrotron 
component and a hard PL with a photon index constrained within $0$-$2$ for the IC component of the nebular emission. 
It is known that the spectrum of the IC component at energies beyond 10~GeV requires  a more complex model. However, 
within our fitting range up to 10 GeV, two PL models are sufficient to characterize the
broad-band spectrum (see  Table~\ref{spec_para_AVE} for the resulting parameters).
More complex models including a ``power law with exponential cutoff'' (PLEC) for the synchrotron component are not
significantly preferred, as a likelihood ratio test indicates that the  improvement is not significant  ($\sim1.3\sigma$). \DHA{It is comforting to see that the sum of these two components (thereafter ``Syn+IC") extrapolated to the $>$100 MeV band agrees within 20\% with that computed from the model of \citet{buehler_gamma-ray_2012}.}

\begin{table*}
	\centering
\caption{Time-averaged spectral properties of the Crab Nebula measured from 60~MeV to 10~GeV.}
\label{spec_para_AVE}

\begin{tabular}{l|ccccc}\hline\hline
Component      & PL $\Gamma$       & PLEC $\Gamma$       & PLEC $E_\mathrm{c}$      & $TS_\mathrm{PLEC}-TS_\mathrm{PL}$~$^\mathrm{a}$ & Integrated Flux~$^\mathrm{b}$      \\
       &        &        & (GeV)      &  & ($10^{-9}$ cm$^{-2}$ s$^{-1}$)      \\ \hline
IC           & 1.415   $\pm$   0.023 & ...               & ...             & ...           & 107       $\pm$        5   \\ 
Synchrotron  & 3.278   $\pm$   0.011 & 3.250    $\pm$    0.026 & 6.6     $\pm$    5.7  & 1.7           & 2534       $\pm$       15  \\ \hline
\end{tabular}

	\raggedright
$^\mathrm{a}$ $\Gamma$ is the photon index. $E_\mathrm{c}$ is the energy of the exponential cutoff. $TS_\mathrm{PLEC}-TS_\mathrm{PL}$ is the difference in test-statistic (TS) between PLEC and PL. \\
$^\mathrm{b}$ The integral fluxes  are based on  PL models. \\

\end{table*}

For evaluation purpose, we repeat the fit with disabled  energy dispersion correction. It turns out that the measured photon index of the synchrotron component becomes steeper by $\sim$0.07, while the difference in the synchrotron photon flux is measured to be  only $\sim$1\%. This indicates that, despite the migration of photon energies, the integrated photon flux in a broad band is approximately conserved. 

\begin{figure}
	\includegraphics[width=9cm]{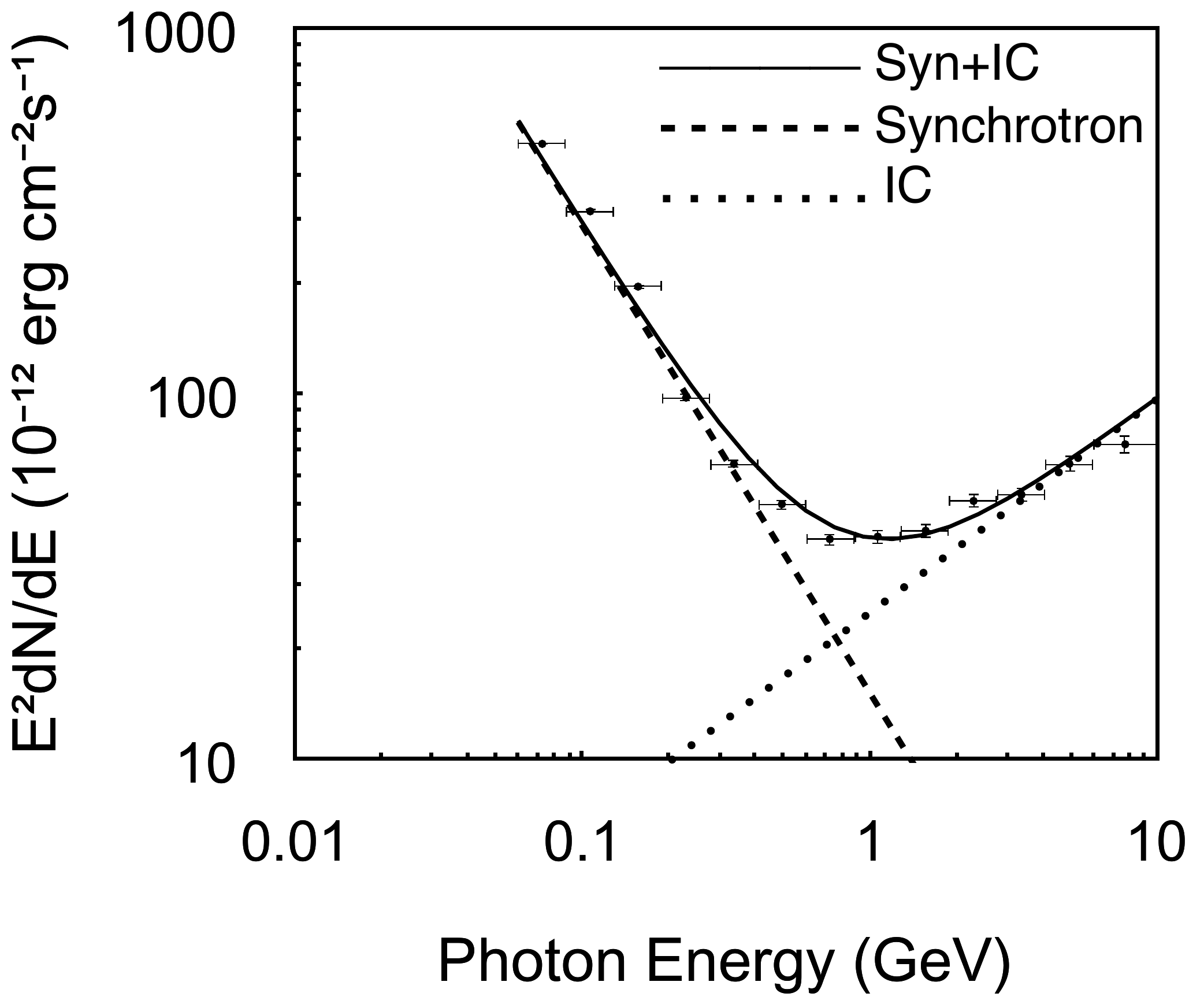}
	\caption{Time-averaged spectral energy distributions (SED) of the Crab Nebula. The solid line and the binned spectrum represent the total Syn+IC emission, while the dashed and dotted lines represent the synchrotron and IC components respectively.\label{SED_AVE}}
\end{figure}

Then, we divide the entire energy band into 13 discrete energy bins (six bins per decade from  
60~MeV to 6~GeV, and a bin between 6~GeV and 10~GeV). 
In the spectral fitting for each bin, we use a single PL component to model the total Syn+IC emission 
 from the Crab Nebula so as to avoid
degeneracies.  Both the photon index and flux normalisation are left free in this procedure.  The measured differential fluxes multiplied by the squared geometrical average energy of
each bin and the corresponding $1\sigma$ uncertainties, as well as the broadband spectral model,  are plotted  in Figure~\ref{SED_AVE}.  The relative systematic uncertainty of the differential flux stemming from disabling the correction for energy dispersion is estimated to be (6--12)\% in 60--600 MeV and (3--6)\% in 600 MeV--10 GeV (see \href{https://fermi.gsfc.nasa.gov/ssc/data/analysis/documentation/Pass8_edisp_usage.html}{Pass 8 Analysis and Energy Dispersion}).

\subsection{Long-term light-curve} \label{subsec:LC}

In order to explore the time-variability of the synchrotron flux, we generate a light-curve for the 60~MeV--600~MeV band. 
In this energy range, we estimate that the IC component only accounts for $<$8\%
of the integrated average flux. It is therefore justified to use a single PL as the model of the total Syn+IC emission for energies
between 60~MeV and 600~MeV.
\DHA{Prior to temporal analyses, we perform an analysis for this energy band with the complete $\sim$10-year data set. The flux normalisation of the isotropic diffuse model is found to be $\approx$1.13 (scaled to the full phase), and the PL spectral index for energy-dependent scaling of the Galactic diffuse model is found to be $\approx$0.018. Since these two parameters
are not expected to noticeably change within 10 years, in analyses for individual temporal segments, we  fix them at the 
$\sim$10-year averages, while the prefactor of the Galactic diffuse model is still left free. }

\begin{figure*}
	\includegraphics[width=18cm]{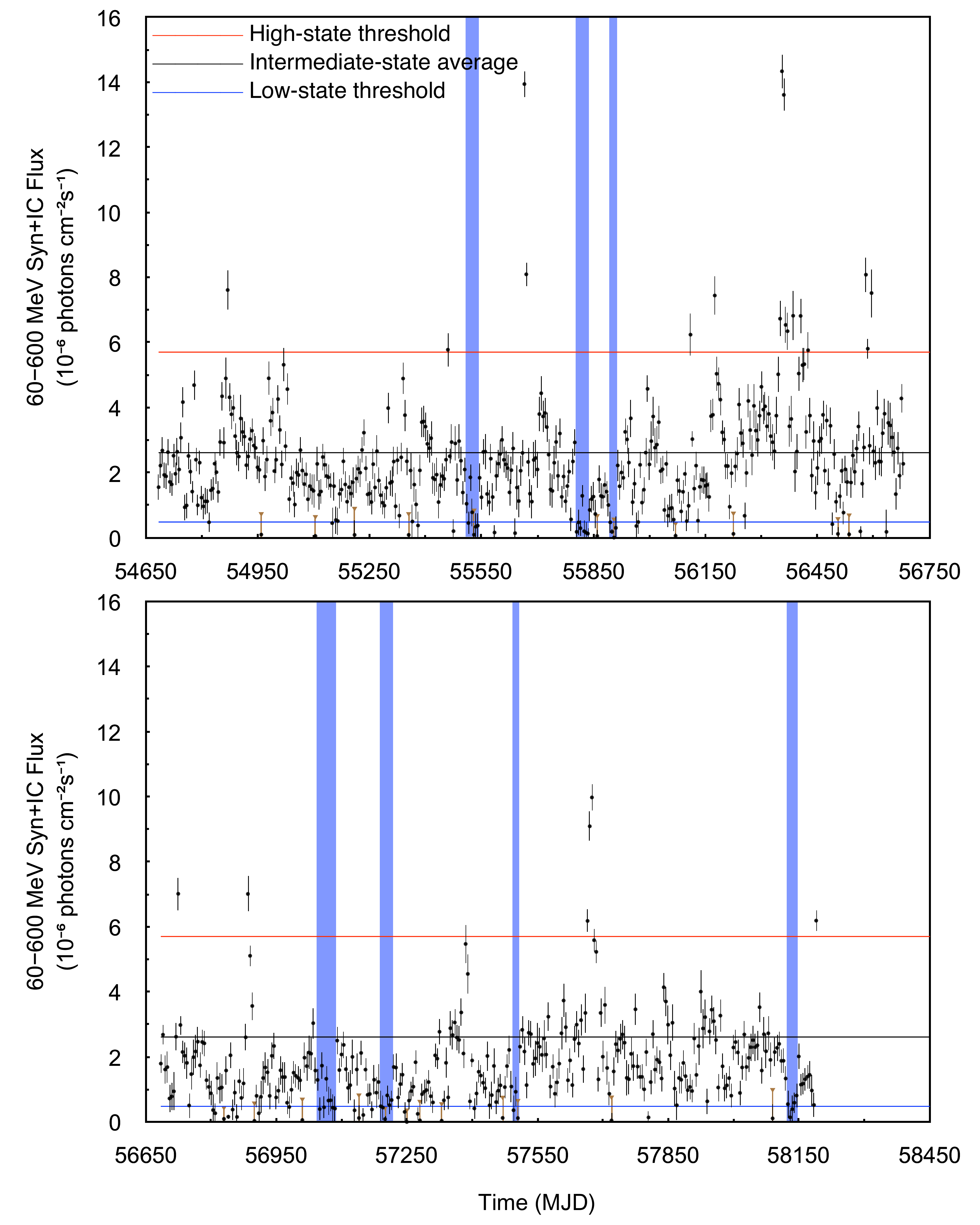}
	
	\caption{Long-term light-curve of the Crab Nebula (the total Syn+IC emission) for the 60~MeV-600~MeV band. The size of each bin is five days. The flux measurements of all bins are plotted as black circles with statistical uncertainties, while the upper limits of a 95\% confidence level (only for the bins with insufficient photon statistics) are plotted as brown triangles. The black horizontal line indicates the intermediate-state average flux, while the red and blue lines respectively indicate the thresholds of the ``high" and ``low" states we define (see the text for detail). \DHA{Blue vertical bands indicate continuous ($\ge$15 d) ``dip`` features which are reported in Table~\ref{epi_quie}.} \label{longterm_lc}} 
\end{figure*}

For the binning of the light-curve we choose a time interval of 5~days which strikes 
a compromise between time-resolution and statistical uncertainties. 
The average photon detection rate from the Crab Nebula during the off-pulse interval chosen here is approximately 100 photons per day. 
In general, the statistical uncertainties of fluxes are conspicuously greater than the photon
shot-noise, reflecting a small signal-to-noise ratio. For those time intervals with insufficient photon
statistics ($\lesssim$4 photons per day), we also place upper limits of a 95\% confidence level 
on the nebular flux. 
The resulting light-curve is shown in Fig.~\ref{longterm_lc}. 

The analysis confirms the finding of previous studies that  the Crab Nebula experiences a series of flares, including those reported by \citet{buehler_gamma-ray_2012,mayer_rapid_2013,Striani2013}, ATELs~\#8519 (Jan-2016, around MJD 57400) and \#9586 (Oct-2016, around MJD 57700). The light-curve is however not well-characterized by a constant flux state 
superimposed by flaring activity with a small duty-cycle, resembling flicker noise. 
For the first time, we find that the flux occasionally drops well-below the average flux value.

\begin{figure}
	\includegraphics[width=9cm]{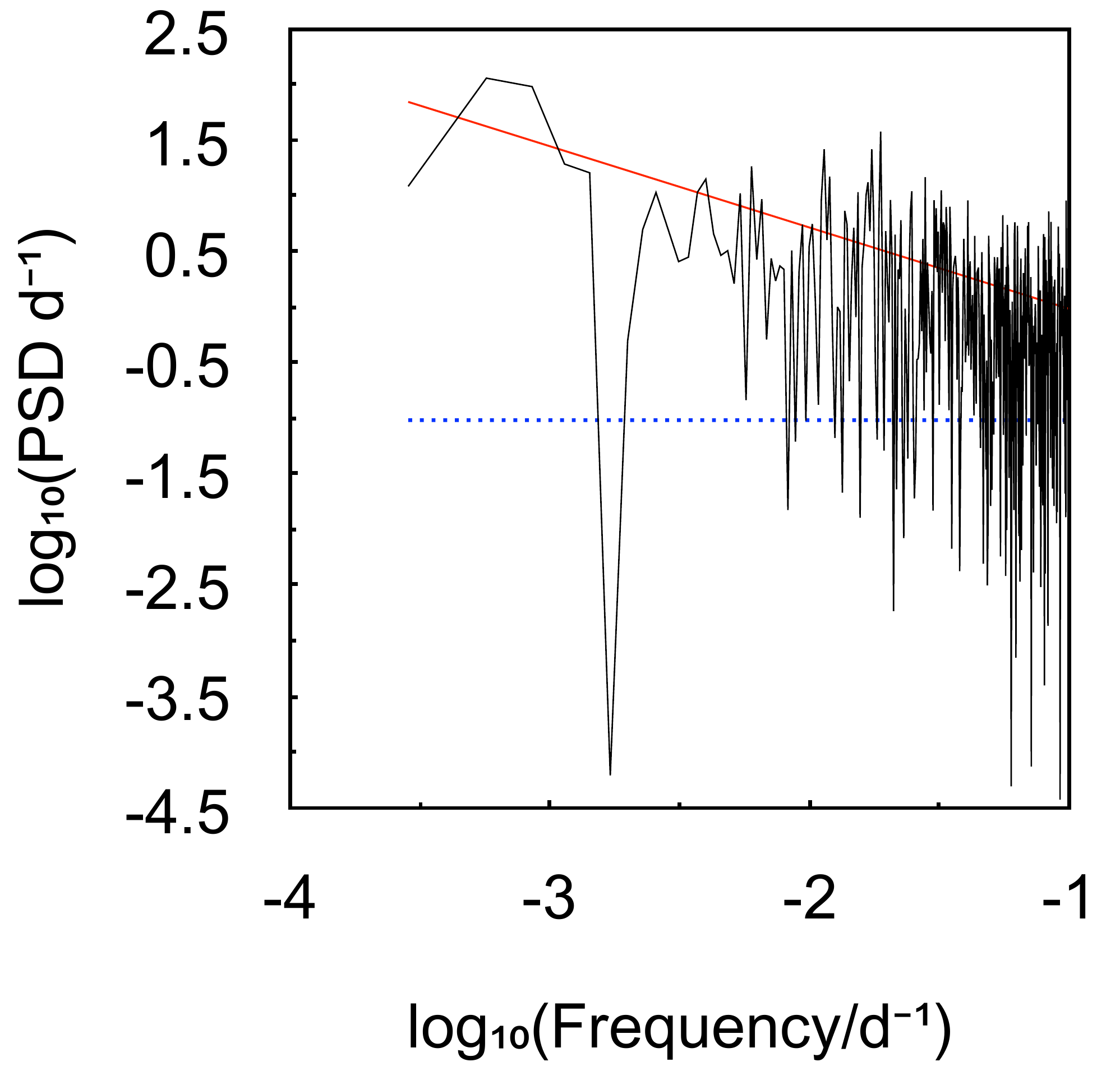}
	
	\caption{Periodogram  obtained from the long-term light-curve of the Crab Nebula. The PSD is normalized to fractional variance per frequency unit. The  red-solid curves in the periodogram respectively indicate the best-fit PL (whose index is $0.73\pm0.12$). The blue-dotted line indicates the white-noise PSD of a control light-curve.  \label{PSD}}
\end{figure}

This impression is confirmed when investigating the light-curve in the 
frequency domain. The periodogram (Figure~\ref{PSD}) is determined via a discrete
Fourier-transformation (DFT) of the real-valued  light-curve normalized to the
average flux.
The power-spectral density (PSD) is calculated from the complex valued coefficients 
of the DFT. The resulting PSD is characterized by a smooth PL such that 
$PSD(f)=(0.18\pm0.08)\times(f/d^{-1})^{-0.73\pm0.12}$. 

\begin{figure}
	\includegraphics[width=9cm]{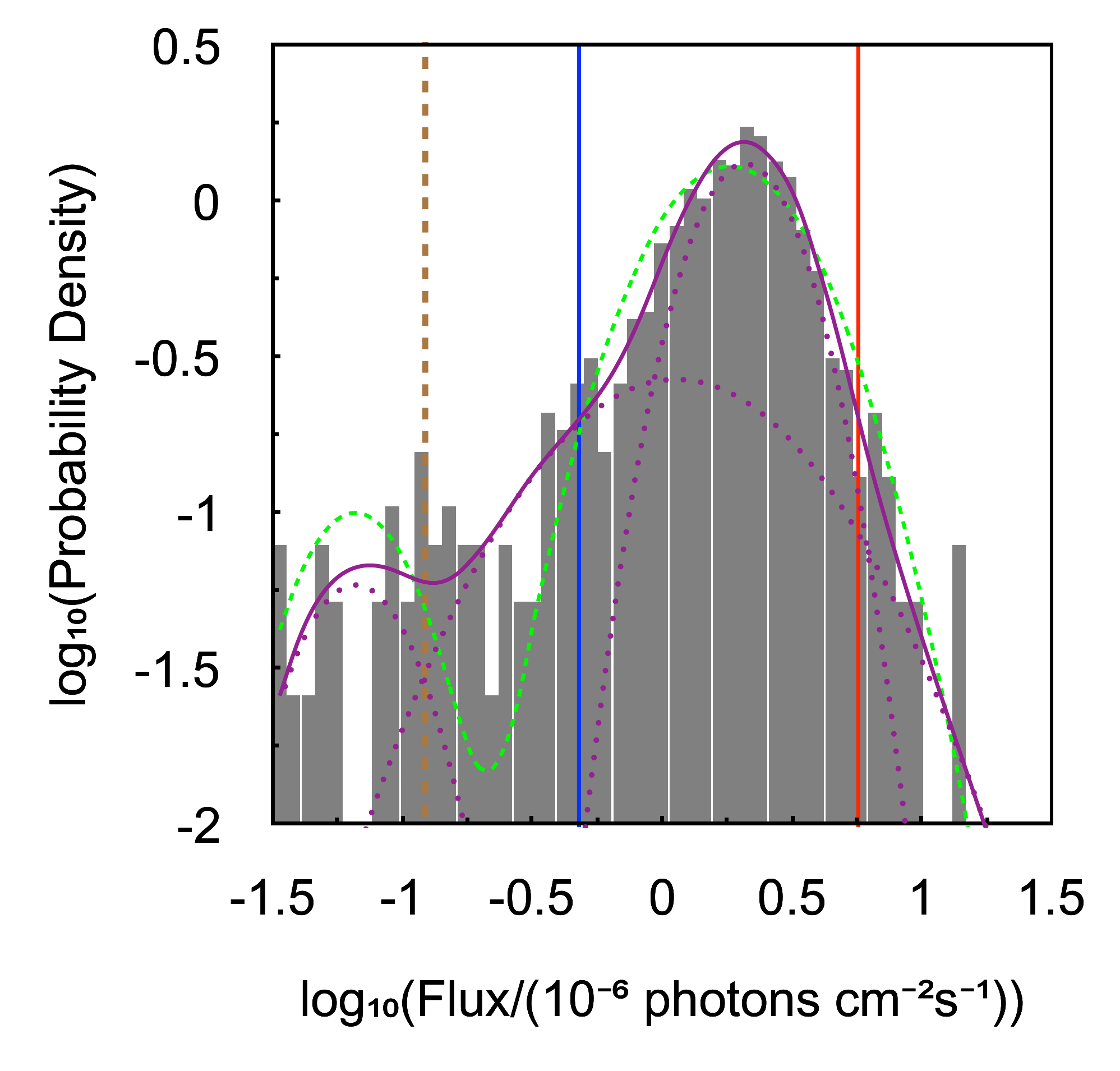}
	
	\caption{Probability density function (PDF)  obtained from the long-term light-curve of the Crab Nebula. The histogram is normalised in a way such that the integration of the probability density over the $\mathrm{log_{10}(Flux)}$ is 1. The double and triple log-normal models fit to the PDF are overlaid as green-dashed and purple-solid curves respectively. Their lowest-flux components model the shot-noise limited distribution of the extrapolated IC flux. The three  components of the triple log-normal model are overlaid as purple-dotted curves. The blue and red vertical lines indicate the threshold of the ``low" and ``high" states respectively. The brown-dashed vertical line indicates the flux sensitivity corresponding to a detection significance of $\sim3\sigma$ and a photon count of $\sim$20 in a 5-day interval.  \label{PDF}}
\end{figure}

The histogram of the flux measurements (Figure~\ref{PDF}) can be described by the superposition of  
two log-normal  distributions (see Table~\ref{pdf_para} for the best-fit paramaters). The component A represents  the extrapolated IC flux fluctuating below/around the detection threshold, with a relative normalisation left free to vary in a Poissonian log-likelihood fit of the histogram. The component B characterises the variable synchrotron emission. This model is preferred over a single log-normal distribution by $\sim13\sigma$, indicating the presence of at least two different flux states. 
Based upon this two-component model, we set the threshold of the ``low" state at 4.8$\times10^{-7}$~cm$^{-2}$ s$^{-1}$, so that the extrapolated tail of the component B below this threshold predicts only less than one-fourth of the observed low-state bins to be contaminated by the component B. The threshold of the ``high" state is set at $5.7\times10^{-6}$~cm$^{-2}$ s$^{-1}$ so that only the top 23 bins are included in the high state. In \S\ref{subsec:simulations}, by using simulated light-curves, we confirm that a two-component model is necessary and sufficient for reproducing the continuous low-flux episodes observed (Figure~\ref{quies_window}).

\begin{table*}
	\centering
\caption{PDF models fit to the histogram of the flux measurements.}
\label{pdf_para}

\begin{tabular}{l|l|ccccc}\hline\hline
Model             & Component & $N_0$~$^\mathrm{a}$        & $F_{maxPD}$~$^\mathrm{b}$       & $\sigma_{lnF}$~$^\mathrm{c}$       & $\Delta{TS}$~$^\mathrm{d}$    & $\Delta{dof}$~$^\mathrm{e}$ \\
                  &           &      (\%)             & ($10^{-6}$~cm$^{-2}$ s$^{-1}$)                      &                                    &       &     \\ \hline
Single log-normal & ...       & 100                 & 1.56             $\pm$ 0.06  & 0.94                        $\pm$ 0.03  & 0     & 0   \\ \hline
Double log-normal & A~$^\mathrm{f}$         & 5.6         $\pm$ 1.0  & 0.066             (fixed) & 0.51                         (fixed) & 180.4 & 1   \\
                  & B         & 94.4        $\pm$ 1.0  & 1.82             $\pm$ 0.05  & 0.67                        $\pm$ 0.02  &       &     \\ \hline
Triple log-normal & X~$^\mathrm{f}$         & 3.3         $\pm$ 1.0  & 0.066             (fixed) & 0.51                         (fixed) & 233.6 & 4   \\
                  & Y         & 31.2        $\pm$ 13.0 & 1.14             $\pm$ 0.33  & 1.07                        $\pm$ 0.18  &       &     \\
                  & Z         & 65.5        $\pm$ 13.9 & 2.10             $\pm$ 0.08  & 0.45                        $\pm$ 0.06  &       &    \\ \hline
\end{tabular}

	\raggedright
$^\mathrm{a}$ The normalisation for scaling a component. In each model, the sum of normalisations of all components must be 1.\\
$^\mathrm{b}$ The flux corresponding to the maximum probability density. It is mathematically equivalent to the exponential of the mean of the flux's natural logarithm.\\
$^\mathrm{c}$ The standard deviation of the flux's natural logarithm.\\
$^\mathrm{d}$ The natural logarithm of the square of the likelihood ratio of a model compared to the single log-normal. The likelihood function is for a Poisson distribution.\\
$^\mathrm{e}$ The number of additional parameters (the extra degrees of freedom) of a model compared to the single log-normal.\\
$^\mathrm{f}$ Components which model the shot-noise fluctuations of the extrapolated IC flux. All of their parameters except normalisations are fixed at values estimated from the photon statistics.\\

\end{table*}

After introducing a third log-normal component, the fitting is further improved by $>6\sigma$.
The component X represents  the extrapolated IC flux and it accounts for the bottom $(3.3\pm1.0)\%$ of measurements. This  corresponds to an expected
$23\pm7$ out of the 68 observed low-state bins. The  strongly variable component Y spans from the low flux state to the highest flux state. The component Z is mostly confined within intermediate flux states. In \S\ref{subsec:spectra_states}, we infer the relative contributions of the synchrotron nebula and IC nebula during the low state, based on spectral analyses.

\DHA{We proceed to perform an analysis with 60--600 MeV data excluding the high- and low-state bins we define. In this way, the intermediate-state average flux is determined to be $(2.61\pm0.02_{stat}\pm0.20_{sys})\times10^{-6}$~cm$^{-2}$~s$^{-1}$, where the systematic uncertainty is determined by altering the prefactor of the Galactic diffuse model by $\pm$10\%. The statistical uncertainty in a 5-day interval is generally more than a double of this systematic uncertainty. The low-state threshold we set is 18.4\% of this intermediate-state average flux.}

\subsection{Systematic effects on the variability}
\label{subsec:systematic}

The instrument, 
its calibration and data analysis contribute  various systematic effects
that may lead to variability in excess of the limiting photon shot-noise.
Similar to the approach presented by
\citet{ackermann_fermi_2012}, we use a data-driven method to investigate systematic 
effects and the stability of the light-curve. 
Fortunately, we can use the Crab pulsar itself to 
establish an estimate of the instrumental variability.

We select data collected during the phase around the highest pulse peak 
(0--0.02,  \& 0.97--1; recall that the phase convention in
\citet{buehler_gamma-ray_2012} is adopted).
Then, from the total Crab flux of each bin, we subtract the nebular flux which 
is measured at the same bin and 
scaled to match the phase interval covering 5\% of the total phase. 
The resulting light-curve for the pulsar emission is based on a photon count
statistics that matches the off-pulse light-curve, making it suitable to be used as a control light-curve.

This control light-curve displays a fractional root-mean-square (RMS) 
variability of $14\%$ with a PSD that is
close to white noise. The resulting PSD can be readily compared with the one
measured from the off-pulse emission (see~Fig.~\ref{PSD}) with a fractional RMS variability of $76\%$. The control light-curve shows some excess noise when compared with the expected fractional variability for
shot-noise only which should be $\approx N_\mathrm{phot}^{-1/2}\approx 4.5\%$
with  the number of photons expected in a 5~d interval ($N_\mathrm{phot}\approx 500$).
We conservatively consider the noise in the control light-curve as an estimate of the instrumental and photon shot noise present in the data. 

The fractional RMS variability of 76\% displayed in the off-pulse light-curve has a significant portion accounted for by  flaring bins. Even if we exclude all bins which are above the intermediate-state average, the fractional RMS variability still remains at a high value of 50\%. This is a strong indication that the flux variations in the nebular light-curve
exceed the  combined systematic and shot noise estimated from the control light-curve at all frequencies (see also Fig.~\ref{PSD}). 

A limitation of the control light-curve is that the 60 MeV--600 MeV spectrum of the Crab pulsar is  harder than that of the nebula \citep{buehler_gamma-ray_2012}. Any  energy-dependent systematic uncertainties of \emph{Fermi} LAT would therefore have  different
impacts on the nebular and pulsar light-curves. As an additional check, we compute the exposure within $1^\circ$ from the Crab for each 5-day bin of the light-curve, assuming a photon index of 3.3. Based upon this study, we \DHA{have no evidence} that the variability is related to fluctuations in the exposure. 

Transient effects due to the relative position of the Sun or the Moon to the Crab Nebula could affect the light-curve. An  
excess of the solar or lunar $\gamma$-ray emission  could lead to an apparent deficit in the computed Crab flux. 
After checking the history of the Sun's position, we do not see \DHA{a causality} between the observed ``dip" features and solar encounters/approaches. 
The lunar encounters/approaches should be comparatively less of an issue, because the $\gamma$-ray emission from the Moon is much less extended
(its radius of $\gamma$-ray extension is only $\lesssim0.5^\circ$) and the Moon remains closer than 7 degrees to the Crab Nebula for 
only $<$1 day (shorter than one-fifth of the bin size) in every sidereal period of 27.3 days. 
The periodogram of the nebular light-curve reveals no distinct modulation at the lunar sidereal period or its harmonics.

Furthermore, the impact of the migration of photon energies on the nebular light-curve leads to additional systematic effects.
While disabling the energy dispersion correction leads to noticeable mis-measurements in the photon index, 
the integrated photon flux in a decade of energy range is expected to remain constant. 
The resulting estimated relative systematic uncertainty on  the photon flux 
($\sim$1\% of the flux, as evaluated in \S\ref{subsec:TimeAveSpec}) is not important when compared to the dominating statistical uncertainty in a 5 d interval.

\subsection{Transitions between low-flux and intermediate states}

\begin{figure*}
	\includegraphics[width=18cm]{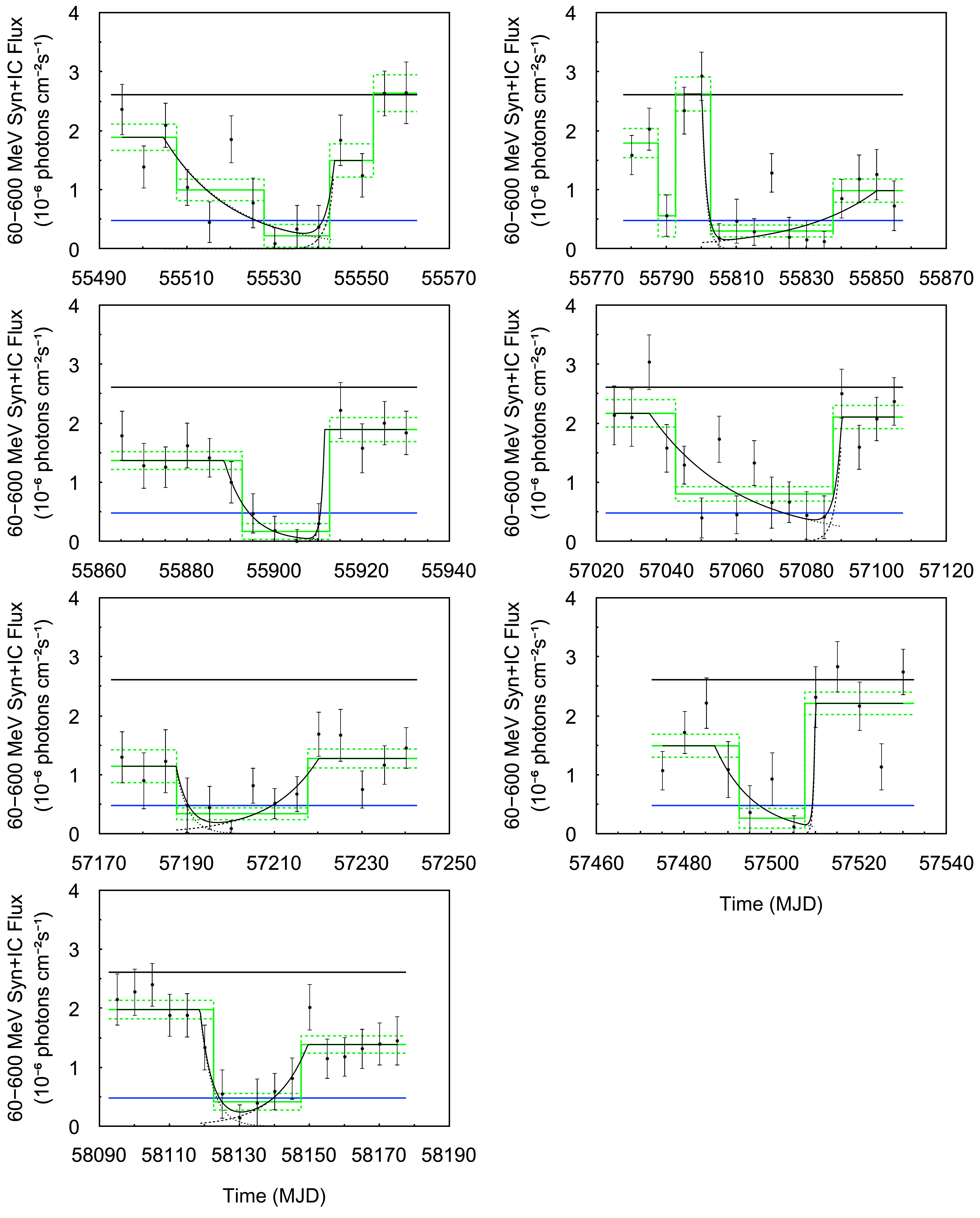}
	\caption{Seven analysis windows covering continuous episodes of low-flux which are tabulated in Table~\ref{epi_quie}. The uniform distribution fit to the bins of each Bayesian block (solid line) and its $1\sigma$ uncertainty (dashed line)  are indicated in green. 
	The  function fit to each segment of the light-curve, as well as its two exponential terms, is plotted as black curves (see the text for detail). The black and blue horizontal lines respectively indicate the intermediate-state average flux and the threshold of the ``low" state we define.\label{quies_window}}
\end{figure*}

We identify seven episodes of continuous low-flux 
where the 60~MeV--600~MeV Syn+IC flux remains as low as 18.4\% of the intermediate-state average for at least half a month.
We apply the Bayesian block algorithm \citep{Scargle_BB_2013} on the seven
analysis windows covering these episodes (Figure~\ref{quies_window}) to
identify different flux states. In turn, we quantify the transitional timescales between them by fitting composite functions to individual 5-day bins in segments of the light-curve. 

We report the time range covered by the lowest 1 or 2 successive blocks of each window as a  low-flux episode. 
The fit range we choose for each window includes the  low-flux episode as well as its preceding and following blocks. The function we fit starts 
and ends with two constant fluxes which are respectively equal to the local averages within the preceding and following blocks of a  low-flux episode. The free
parameters of the fit include the starting and stopping times of the low-flux episode where   
the flux varies as a sum of an exponential decay term and an exponential growth term. The predicted flux must be continuous in the whole fit range. There are in total four free parameters in the fit: in addition to the starting and stopping times, we estimate
the halving time of the decay term as well as the doubling time of the growth term. 

The best segmentations with a false positive rate of 0.07, as well as the functions fit to segments of the light-curve,  are overlaid in Figure~\ref{quies_window}. The information about the seven episodes and the timescales of transitions are tabulated in Table~\ref{epi_quie}. As a cross-check, we repeat the fits with two additional free parameters: the constant flux before decay and that after growth. We obtain  consistent results. The fit results 
constrain the shortest timescales of transitions between low-flux and intermediate states to be $<1.9$ days (95~\% c.l.). 

\begin{table*}
	\centering
\caption{Information about seven episodes of continuous low-flux.}
\label{epi_quie}
\begin{tabular}{lcccc}\hline\hline
Start Time & Duration~$^\mathrm{a}$ & $F_\mathrm{low}$~$^\mathrm{b}$      & $t_{1/2}$~$^\mathrm{c}$      & $t_2$~$^\mathrm{d}$      \\
(MJD)      & (days)   & ($10^{-6}$~cm$^{-2}$ s$^{-1}$)              & (days)       & (days)        \\ \hline
55507.7    & 35       &      0.70$\pm$0.14      &  10.7$^{+4.7}_{-2.8}$	&	$<$5.3\\
55802.7    & 35       &      0.38$\pm$0.11      &  $<$2.2	&	15.5$^{+31.6}_{-6.1}$\\
55892.7    & 20       &      0.24$\pm$0.15      &  3.8$^{+2.9}_{-1.9}$	&	$<$3.3\\
57042.7    & 45       &      0.82$\pm$0.12      &  17.7$^{+4.5}_{-3.3}$	&	$<$4.0\\
57187.7    & 30       &      0.50$\pm$0.13      &  $<$12.6	&	7.5$^{+2.2}_{-3.1}$\\
57492.7    & 15       &      0.47$\pm$0.22      &  6.2$^{+2.6}_{-2.3}$	&	$<$1.9\\
58122.7    & 25       &      0.50$\pm$0.16      &  2.4$^{+2.6}_{-1.1}$	&	6.4$^{+3.4}_{-2.9}$\\ \hline
\end{tabular}

	\raggedright
$^\mathrm{a}$ The start times and durations are determined from Bayesian block segmentations.
 \\
$^\mathrm{b}$ The local average fluxes within the durations. In view of a broad variety of uncertainties, we adopt the unweighted means (instead of the error-weighted means plotted in Figure~\ref{quies_window}) for unbiased calculations.
 \\
$^\mathrm{c}$ Halving times of the exponential decay term. The upper limits are at a 95\% confidence level.
 \\
$^\mathrm{d}$ Doubling times of the exponential growth term. The upper limits are at a 95\% confidence level.
\\

\end{table*}

\subsection{Comparison of the observed light-curve to simulated light-curves} \label{subsec:simulations}

The nebular emission is characterized by a red-noise PSD, dominating above 
instrumental noise at all frequencies sampled. In the time-domain, we have
identified
episodes where the flux of the nebula drops well below the average and remains low for 
several weeks. In order to clarify to what extent these kind of episodes occur randomly, we 
simulate $10^6$ light-curves following the recipe of \citet{Emmanoulopoulos2013} which 
has been implemented in the ``DELightcurveSimulation" package \citep{Connolly2015}. The method
extends on the original approach \citep{tam_production_2012}, where
a method to simulate light-curves with Gaussian distributed flux states and a power-law PSD is introduced. In the
method used here, an arbitrarily shaped probability density function (PDF) for the
flux state can be used. 

The bulk of observed  flux states follows  a log-normal distribution.
However, a noticeable deviation at lower flux states is apparent (see Fig.~\ref{PDF}). 
We simulate therefore a  log-normal PDF (with the same mean and standard deviation as the component B in Table~\ref{pdf_para}) 
in combination with the power-law PSD (Figure~\ref{PSD}). In absence of a  low state,
we can use the simulated light-curves to estimate the probability of appearance of 
similar episodes of low-flux as observed in the data.

We apply the Bayesian block algorithm \citep{Scargle_BB_2013} on our observed light-curve and each simulated light-curve,
with a false positive rate of 0.07. Then, we search for the continuous ``dip" feature, which is defined
as a block or a set of successive blocks fulfilling  two conditions (mimicking the phenomena shown in Figure~\ref{quies_window}): 
\begin{itemize}
  \item The total length is at least 3 bins (15 days).
  \item The local mean (error-weighted) of each included block is below 4.8$\times10^{-7}$~cm$^{-2}$ s$^{-1}$ (the blue line in Fig.~\ref{longterm_lc}~\&~\ref{quies_window}). 
\end{itemize}
Such a dip feature appears in our observed light-curve for a total of 7 times.

Among the simulated light-curves based on the  log-normal PDF,  only a fraction of  $5.4\times10^{-5}$ have  $\ge$7 dips. 
In other words, the expected number of dips in a simulated light-curve is less than that in our observed light-curve at a $>3.8\sigma$ level. 

In order to verify that a PDF with a second, low-flux component is a closer match to the
observed features in the light curve, we simulate again $10^6$ light curves using a double log-normal  distribution (Table~\ref{pdf_para}) in combination with the same PSD. 
For this PDF, the average number of dips in a simulated light-curve is $6.0\pm1.9$, which is  consistent with our observations.

Repeating two chains of simulations with a more complex PSD (curved and with a constant additive term), 
we obtain very similar results, verifying that the exact shape of the model for the PSD is not of importance. 
While a double log-normal PDF is sufficient for a simulation to reproduce the 7 continuous dip features we observe, we recall that the whole histogram of flux measurements suggests a more complicated distribution of flux states (see Figure~\ref{PDF} and Table~\ref{pdf_para}).

\subsection{Spectra in different flux states}\label{subsec:spectra_states}

\begin{figure*}
	\includegraphics[width=18cm]{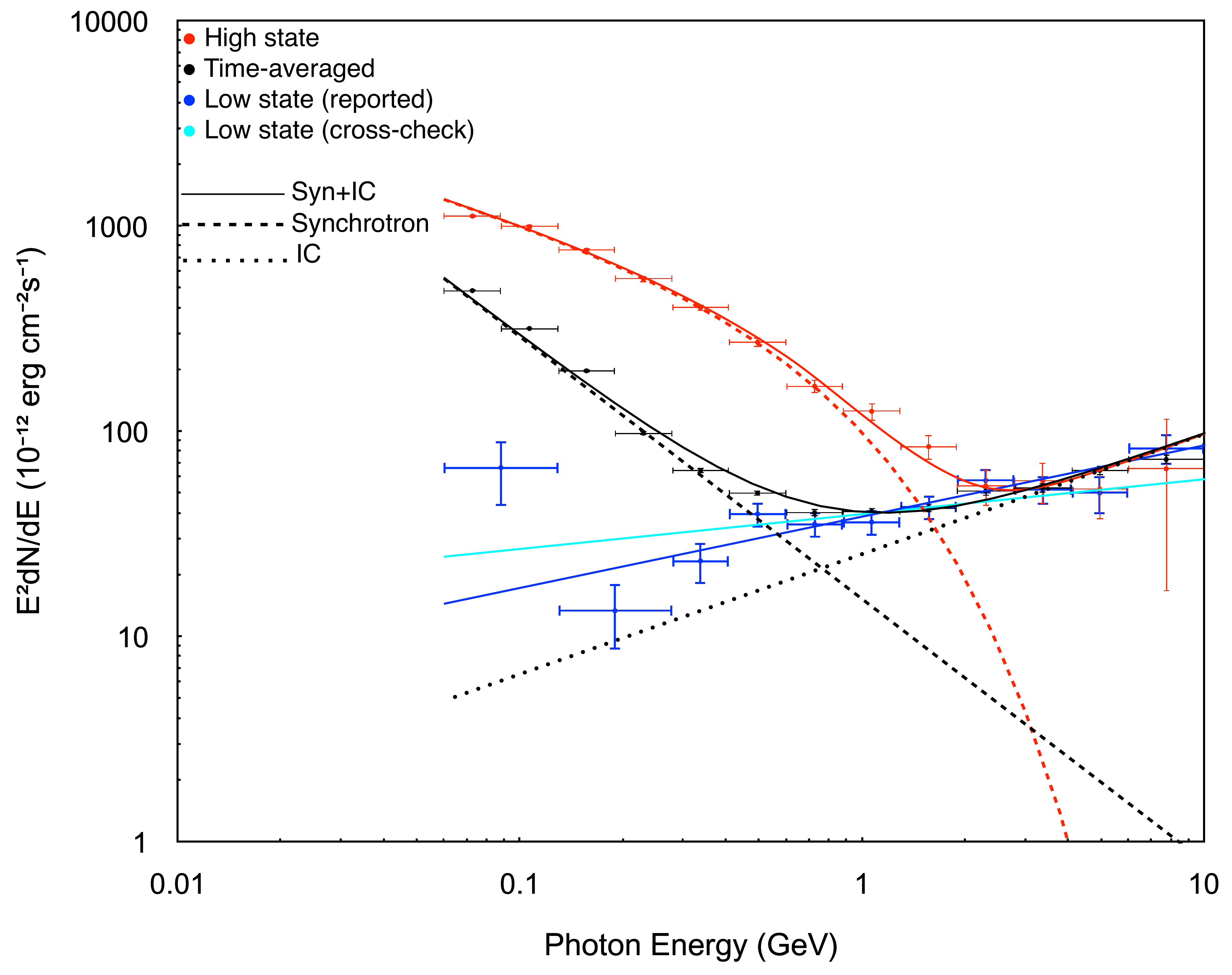}
	\caption{Spectral energy distributions (SED) of the Crab Nebula for different flux states.  The  IC
	component (the dotted line) is determined with the time-averaged spectrum (in black). 
	The spectra of the high and low states (in red and blue respectively) are defined based on the
	60~MeV-600~MeV light-curve of the Crab Nebula (Figure~\ref{longterm_lc}). 
	The solid lines and the binned spectra represent the total Syn+IC emission, while the dashed lines
	represent the synchrotron component. For the low state (solid blue line), 
	the combined Syn+IC spectrum is plotted as a single PL component. The solid cyan line is an alternative low-state spectrum for cross-checking (see the text for detail). \label{SED}}
\end{figure*}

The result of the temporal analysis suggests the existence of a  low-flux state  (see \S\ref{subsec:LC}~\&~\S\ref{subsec:simulations}).
In order to investigate the spectral changes of the nebula during the defined ``high" and ``low" 
states respectively, we sort the 702 bins of the light-curve according to the best-fit photon flux. 
The thresholds of these two flux states have been shown in Figures~\ref{longterm_lc}~\&~\ref{PDF}. We group the top 23 bins above the red line into the high flux state data. 
For the low state, we select the lowest 68 bins below the blue line.  Their accumulated TS is sufficient for us to create a binned spectrum with well-constrained uncertainties.
We repeat the chain of
spectral fittings described in \S\ref{subsec:TimeAveSpec} for the high and low states. The results are plotted in red and blue
respectively in Figure~\ref{SED}. The fit-parameters are tabulated in Table~\ref{spec_para}. 

\begin{table*}
	\centering
\caption{60~MeV--10~GeV spectral properties of the Crab Nebula measured in different flux states.}
\label{spec_para}

\begin{tabular}{l|l|ccccc}\hline\hline
Component   & State~$^\mathrm{b}$   & PL $\Gamma$       & PLEC $\Gamma$       & PLEC $E_\mathrm{c}$      & $TS_\mathrm{PLEC}-TS_\mathrm{PL}$~$^\mathrm{c}$ & Integrated Flux~$^\mathrm{d}$      \\
   &    &        &        & (GeV)      &  & ($10^{-9}$ cm$^{-2}$ s$^{-1}$)      \\ \hline
IC~$^\mathrm{a}$          & Average & 1.415   $\pm$   0.023 & ...               & ...             & ...           & 107       $\pm$        5   \\ \hline
            & High    & 2.830   $\pm$   0.015 & 2.497    $\pm$    0.041 & 0.77    $\pm$    0.10 & 114.1         & 8432       $\pm$       92  \\
Synchrotron & Average & 3.278   $\pm$   0.011 & 3.250    $\pm$    0.026 & 6.6     $\pm$    5.7  & 1.7           & 2534       $\pm$       15  \\
            & Low     & ...             & ...               & ...             & ...           & 370       $\pm$        130 \\ \hline
Syn+IC      & Low     & 1.653   $\pm$   0.048 & ...               & ...             & ...           & 221       $\pm$       22   \\ \hline
\end{tabular}

	\raggedright
$^\mathrm{a}$ The parameters of the IC component are determined from the complete $\sim$10-year data set, and are assumed to remain constant. \\
$^\mathrm{b}$ The high and low states are defined based on the 60~MeV-600~MeV  light-curve of the Crab Nebula (Figure~\ref{longterm_lc}). \\
$^\mathrm{c}$ $\Gamma$ is the photon index. $E_\mathrm{c}$ is the energy of the exponential cutoff. $TS_\mathrm{PLEC}-TS_\mathrm{PL}$ is the difference in test-statistic (TS) between PLEC and PL. \\
$^\mathrm{d}$ For the high state and the average, the integral fluxes of the synchrotron component are based on PLEC and PL models respectively. For the low state, it is the sum over the binned spectrum subtracting the IC component.\\

\end{table*}

In both states, the binned spectra indicate that the differential flux at any energy between
1.9~GeV and 10~GeV remains consistent with the $\sim$10-year average, within the tolerance of 
$1.5\sigma$ uncertainties.
Therefore, we fix the parameters of the IC component at the values determined with the whole
$\sim$10-year data set.

During the high state, the PL index of the synchrotron component is harder than that of the 
$\sim$10-year average spectrum by $\sim24\sigma$, and PLEC is preferred over PL by 
$\sim10.7\sigma$, confirming previous results on the flaring state of the Crab Nebula 
\citep{buehler_gamma-ray_2012,mayer_rapid_2013}.
The differential low-state spectrum (shown in Fig.~\ref{SED} in blue) differs
from the average spectrum too.  During the low state, the energy spectrum of the synchrotron component cannot be well described by PL or PLEC with 
physically reasonable parameters, so we just report the synchrotron flux computed 
directly from the 
binned spectrum, which is $(15\pm5)\%$ of the $\sim$10-year average. 

On the other hand, the 
entire Syn+IC spectrum during the low state can be  fit by a single PL component, \DHA{despite a potential excess in the lowest energy bin  ($\sim2.5\sigma$). Such an outlying bin is measured with 60--129~MeV data which is limited by particularly poor spatial and spectral resolutions of LAT as well as severe inaccuracy of diffuse models.}
This PL has a hard index ($1.65\pm0.05$) and a low integral flux which are quite comparable to those of the $\sim$10-year average IC component.  The $\gamma$-ray luminosity of the IC component is 77\% of the  low-state luminosity of the whole Crab Nebula computed from this model.

We note that 22 out of the 68 low-state bins have their preceding and following bins both $\ge$44\% of the intermediate-state average. They can be considered ``isolated" (i.e., not in pairs or clusters). On average, a mock light-curve simulated with the  log-normal PDF and the power-law PSD (reported in \S\ref{subsec:simulations}) has $16.2\pm4.0$ low-state bins where $2.0\pm1.7$ of them are isolated in the same way. We recall that the  combined systematic and shot noise has a fractional RMS variability of $\sim14\%$ (as evaluated in \S\ref{subsec:systematic}). Also, immediately before/after a low-flux bin, the Crab Nebula is probably in a similar physical state for a while. These entail that some  isolated low-state bins in the observed nebular light-curve could be occasional chance events. Therefore, the numerous discontinuities in our selection of low-state bins could have introduced non-negligible systematic bias in measuring the low-state spectral properties. 

With regards to this issue, we reconstruct an alternative low-state spectrum as a cross-check. In order to investigate the spectrum for clusters of low-flux bins, we group a total of 41 bins of the seven continuous low-flux episodes (Table~\ref{epi_quie}) into this alternative low-state, and the obtained result is overlaid in Figure~\ref{SED} as well. It turns out that the two low-state spectra have very similar integrated fluxes and photon indices.  

\section{Summary \& conclusion}

\subsection{Summary of main results}

\paragraph{Variability and low-flux state of the synchrotron nebula}
The long-term light-curve of the gamma-ray emission from the Crab Nebula in the
energy range between 60~MeV
and 600~MeV  has been extracted from $\sim$10 years of observation with the
\emph{Fermi} LAT instrument. On average, $>$92\% of the integrated flux is accounted for by
synchrotron radiation. The light-curve 
shows pronounced variability, with a relative standard deviation equal to 76\%.
As demonstrated with a control light-curve from a phase-gated part of the pulsar emission, we estimate that less than 2~\% of the measured
variability could be related to instrumental or systematic effects. 
The periodogram follows a  PL with an index of $0.73\pm0.12$, indicating
the presence of flicker-noise in the entire frequency range covered by the
observations. In the observed light-curve, we identify at least 7  episodes  during which
the source flux drops below  18.4\% of the intermediate-state average. Using Bayesian
blocks, we characterize these episodes to last between 5 and 35 days. We have used
simulated light curves to estimate the probability of chance appearance of these
episodes for a variable source characterized by a single log-normal
distribution of flux states and a PSD with the same spectral shape as found for the
observed light curve. We infer a probability of $\sim5.4\times10^{-5}$ to have the number of continuous ($\ge$15 d) low-flux episodes detected in a simulation greater than or equal to that in our observed light-curve.  A superposition of  two log-normal  distributions is sufficient for a simulation to reproduce the 7 continuous dip features we observe. On the other hand, a PDF model containing three log-normal components is statistically favored to describe the entire histogram of flux measurements. 
\paragraph{Energy spectrum during different flux states}
The energy spectra have been extracted in three  flux intervals respectively.
The binned spectra in the energy range from 2 GeV to 10 GeV implies that the state transitions do not lead to any 
noticeable change in the IC component up to  10 GeV. After all, the IC component is intrinsically steady during the lifetime of the Crab Nebula, because the responsible low-energy electrons fill a large volume with a cooling timescale exceeding the age of the nebula. 
We confirm the general trend of a hardening and curvature of the synchrotron spectrum in the high-flux state, which is discussed in \citet{buehler_gamma-ray_2012} 
and \citet{mayer_rapid_2013}. For the first time, we reconstruct the energy spectrum in the newly found low-flux state. 
The energy spectrum in the low-flux state and at energies below 2 GeV is roughly consistent  with an extrapolation of  
the IC component of the nebula emission towards lower energies.

Notably, the fitting for the IC component is dominated by the  $>$2~GeV data, leading to a large uncertainty in its extrapolated flux below  600 MeV. Also, we found an energy-dependent spatial extension of the IC nebula, where the size shrinks as the photon energy  increases \citep{Yeung_Morphology_2019}. The extrapolated extension size (the 68\% containment radius) at 100 MeV is as large as $0.1^\circ$. However, we model the IC component as a point source in this work, which only accounts for a core part of the IC nebula. Therefore, the extrapolated IC flux could have been  underestimated while the measured low-state spectrum provides an indication to the actual IC nebula emission.

We therefore conclude from the characterization of the variability and the spectral analysis that the synchrotron nebular emission between 60~MeV and 
600~MeV drops well below the average flux on time-scales of several days and remains in a low state for several weeks. 
During these episodes of low-state emission, the predomination of the nebular energy spectrum by the IC emission  
demonstrates that \DHA{the high-energy part of the synchrotron nebula is dramatically depressed on a short time-scale. }

We consider in the following a possible interpretation of a compact emission region which satisfies the requirement that
the emission region is causally connected within the variability time-scale found during the transition phase of less than $2$ days. This compact region
would be the origin of the bulk of the observed emission such that it would explain simultaneously the rapid dimming of the entire emission as well as the 
low-state spectrum which is apparently dominated by the constant flux of the IC nebula. Possible alternative explanations based upon 
variability of the entire nebula need to circumvent 
the argument of causal connection.

We focus here on the 
well-known inner knot observed near the pulsar's position \citep{hester_wfpc2_1995} as a possible candidate. 

\subsection{Interpretation as synchrotron emission from the inner knot}
    With the shortest timescales of transitions between low-flux and intermediate states constrained to be $<2$  days (95\,\% c.l.), 
    we infer that at least 75\% of the $>10^8$~eV emission of the so-called synchrotron nebula originate from a  compact region with an extension  limited by the light crossing time to be $ct_\mathrm{var}\approx 4.2~\mathrm{mpc}~t_\mathrm{var}/(5~\mathrm{d})$ 
    which corresponds to an angular diameter (at a distance of 2.2~kpc) of $\theta\approx\ang[angle-symbol-over-decimal]{;;0.4}~t_\mathrm{var}/(5~\mathrm{d})$. 
   The time-scale of variability and the inferred angular extension of $\ang[angle-symbol-over-decimal]{;;0.4}~t_\mathrm{var}/(5~\mathrm{d})$ is consistent with the finding of \citet{rudy_characterization_2015}, where the tangential FWHM of the knot was observed to be $0.3''-0.35''$.
     The result of our analysis of the variability therefore strengthens the interpretation that the high-energy part of the synchrotron emission is produced in the inner knot of the Crab nebula as put forward by \citet{komissarov_origin_2011}.
    
     The inner knot has also been found to show variability in the optical and X-ray band \citep{rudy_characterization_2015} with correlations of the knot's morphology/position with its gamma-ray flux that are 
    similar to the expectations of models of the termination shock \citep{lyutikov_inner_2016}. Further multi-wavelength observations of the inner knot during a phase of 60 MeV--600 MeV low-state would be essential to confirm the proposed scenario.


\begin{acknowledgements}
PKHY acknowledges the support of the DFG under the research grant 	HO 3305/4-1. We greatly appreciate M. Kerr for providing the ephemeris of the Crab pulsar for phased analysis. PKHY thanks H.-F. Yu for useful discussions. We thank the two anonymous referees for very useful comments which helped to improve the manuscript.
\end{acknowledgements}

\vspace{5mm}

\bibliographystyle{aa}



\end{document}